\documentclass[
    aps,%
    12pt,%
    final,%
    notitlepage,%
    oneside,%
    onecolumn,%
    nobibnotes,%
    nofootinbib,%
    superscriptaddress,%
    noshowpacs,%
    centertags]%
    {revtex4-2}

    \usepackage{amsmath}
    \usepackage{graphicx}
    \usepackage{array}
    \usepackage[colorlinks=true, allcolors=blue]{hyperref}
    \usepackage[T2A, T1]{fontenc}
    \usepackage[utf8x]{inputenc}
    \usepackage[english]{babel}
    \usepackage[export]{adjustbox}%
    \usepackage{booktabs}
    
    \usepackage{amsmath}
    \usepackage{graphicx}
    \usepackage{subcaption}
    \usepackage{float}
    \usepackage{array}
    \usepackage[colorlinks=true, allcolors=blue]{hyperref}
    \begin{document}

    \title{A Methodology for Developing Foundational Transformer Models in Collider Physics Analysis}
    
    \author{\firstname{E.}~\surname{Abasov}}
    \affiliation{Skobeltsyn Institute of Nuclear Physics of Lomonosov Moscow State University (SINP MSU), 1(2), Leninskie gory, GSP-1, Moscow 119991, Russian Federation}
    \author{\firstname{L.}~\surname{Dudko}}
    \affiliation{Skobeltsyn Institute of Nuclear Physics of Lomonosov Moscow State University (SINP MSU), 1(2), Leninskie gory, GSP-1, Moscow 119991, Russian Federation}
    \author{\firstname{E.}~\surname{Iudin}}
    \affiliation{Skobeltsyn Institute of Nuclear Physics of Lomonosov Moscow State University (SINP MSU), 1(2), Leninskie gory, GSP-1, Moscow 119991, Russian Federation}
    \author{\firstname{A.}~\surname{Markina}}
    \affiliation{Skobeltsyn Institute of Nuclear Physics of Lomonosov Moscow State University (SINP MSU), 1(2), Leninskie gory, GSP-1, Moscow 119991, Russian Federation}
    \author{\firstname{P.}~\surname{Volkov}}
    \affiliation{Skobeltsyn Institute of Nuclear Physics of Lomonosov Moscow State University (SINP MSU), 1(2), Leninskie gory, GSP-1, Moscow 119991, Russian Federation}
    \author{\firstname{M.}~\surname{Perfilov}}
    \affiliation{Skobeltsyn Institute of Nuclear Physics of Lomonosov Moscow State University (SINP MSU), 1(2), Leninskie gory, GSP-1, Moscow 119991, Russian Federation}
    \author{\firstname{A.}~\surname{Zaborenko}}
    \email{zaborenko.ad18@physics.msu.ru}
    \affiliation{Skobeltsyn Institute of Nuclear Physics of Lomonosov Moscow State University (SINP MSU), 1(2), Leninskie gory, GSP-1, Moscow 119991, Russian Federation}
    
    \begin{abstract}
    We present a methodology for training foundational transformer models capable of processing collider data with diverse kinematic signatures. 
    Our universal foundation model is designed for simultaneous analysis of all processes involving from one to four top-quarks production with their corresponding background processes. The approach employs multi-task pre-training on combined datasets of simulated events, enabling the model to capture the full spectrum of interaction physics while extracting universal patterns across different final states prior to task-specific fine-tuning. This unified architecture eliminates the need for separate analysis frameworks for different final signatures and specific tasks.
    
    The transformer-based pre-training strategy explicitly preserves unique interaction patterns through adaptive attention mechanisms while establishing cross-process correlations. We plan to demonstrate how this architecture maintains sensitivity to rare high-multiplicity topologies (3t and 4t) without compromising performance on conventional channels ($\rm t\bar t$, tX, $\rm t\bar t H$), effectively bridging the gap between disparate analysis paradigms in collider physics.
    \end{abstract}
    \maketitle
    \section{Foundational Models}
    
    In the fields of computer vision~\cite{transformer_cv} and natural language processing~\cite{transformer_nlp}, foundational models pre-trained on large and diverse datasets are widely used. Such training allows models to uncover data patterns useful for numerous downstream tasks.
    
    In High Energy Physics (HEP), this approach shows particular promise for event classification. A foundational model capturing general physical patterns (e.g., energy-momentum correlations, jet substructure relationships) could enhance the accuracy of specialized analyses through transfer learning.
    
    The study investigates:
    \begin{itemize}
    \item Architectural adaptations for collider data (input space formulation, numerical embeddings)
    \item Comparative evaluation of pre-training techniques (masked reconstruction vs contrastive learning)
    \item Transfer learning capabilities through downstream regression and classification tasks
    \end{itemize}
    
\begin{figure}[h]
    \label{fig:ae_scheme}
     \includegraphics[width=0.5\linewidth]{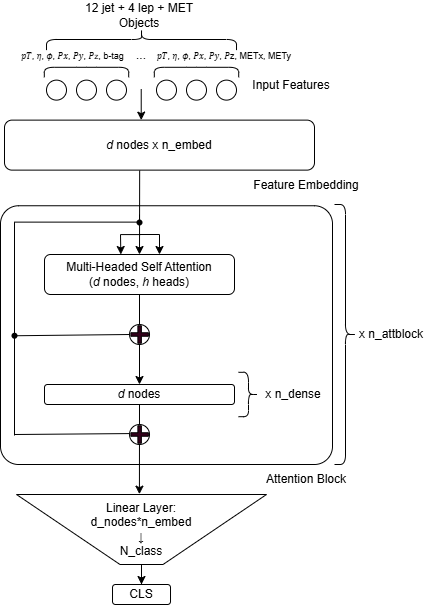}
    \caption{Graphical representation of the Transformer model. Input features are embedded through a LinearEmbedding~\cite{gorishniy2022embeddings} layer, then passed through several Attention Blocks employing Multi-Headed Self-Attention. The output of the Transformer is flattened and passed through the final Linear Layer to convert its dimension to $N_{class}$ for classification. This approach can be used to convert the output of the model into any number of dimensions for regression or classification tasks.}
    \label{fig:transformer_scheme}
    \end{figure}

    For analyzing complex 3- and 4-top-quark production processes, we developed a large neural network (NN) architecture (Fig.~\ref{fig:transformer_scheme}), pre-trained on kinematically distinct production processes sharing underlying Standard Model (SM) physics. This model bridges the gap between specialized networks (trained on specific kinematic regimes) and broad foundational models capable of pattern recognition across entire HEP domains.
    
    \section{Unified input variable space}
    
\begin{figure}[H]
        \centering
        \includegraphics[width=0.3\textwidth]{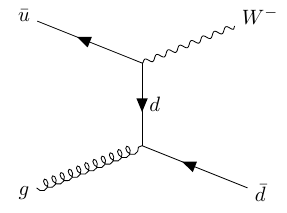} \hfill
        \includegraphics[width=0.3\textwidth]{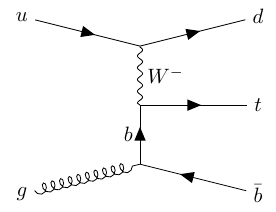} \hfill
        \includegraphics[width=0.3\textwidth]{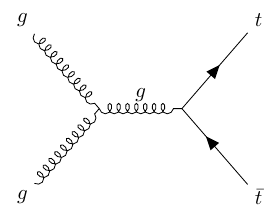} \\[1em]
        
        \hspace*{\fill} 
        \includegraphics[width=0.3\textwidth]{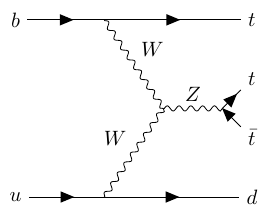} \hfill
        \includegraphics[width=0.3\textwidth]{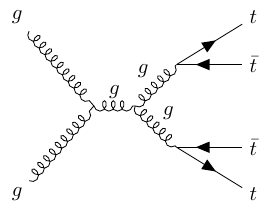}
        \hspace*{\fill} 
        
        \caption{Example Feynman diagrams for process groups with different top-quark counts.}
        \label{fig:t_q_diag}
    \end{figure}

    For testing these approaches, a dataset of simulated collider events was used, comprising five process groups categorized by top-quark multiplicity (Fig.~\ref{fig:t_q_diag}). Event generation utilized MadGraph5~\cite{Alwall:2014hca} and CompHEP~\cite{CompHEP:2004qpa,Pukhov:1999gg} packages with the number of events for each process scaled according to available computational resources, while detector response simulation employed in DELPHES~\cite{Selvaggi:2014mya} for the CMS detector (13 TeV) ~\cite{CMS_detector, CMS_card}.
    
    The dataset is categorized into five classes based on the number of top quarks in the final state:
    
    \begin{itemize}
        \item \textbf{Zero-top class:}  
        \begin{itemize}
            \item $W^+W^-$ production (diboson)  
            \item $W$+jets production (inclusive $W$ boson with jets)  
        \end{itemize}
        
        \item \textbf{Single-top class:}  
        \begin{itemize}
            \item Single-top production via $t$-channel (anti-top and top quarks)  
            \item Single-top production in association with a $W$ boson ($tW$-channel for anti-top and top quarks)  
        \end{itemize}
        
        \item \textbf{Two-top class:}  
        \begin{itemize}
            \item $t\bar{t}$ production (top quark pair) in dileptonic and semileptonic decay channels  
        \end{itemize}
        
        \item \textbf{Three-top class:}  
        \begin{itemize}
            \item Standard Model (SM) processes: $TTTJ$ (triple-top + jet) and $TTTW$ (triple-top + $W$ boson)  
        \end{itemize}
        
        \item \textbf{Four-top class:}  
        \begin{itemize}
            \item SM four-top-quark production ($TTTT$)  
        \end{itemize}
    \end{itemize}
    
    \begin{table}[H]
    \centering
    \begin{tabular}{lrrrr}
    \hline
    Class & Count & Percentage (\%) & Class Weight \\
    \hline
    Zero-top & 1,787,876 & 21.81 & 0.917 \\
    Single-top & 2,174,999 & 26.53 & 0.754 \\
    Two-top & 3,329,795 & 40.62 & 0.492 \\
    Three-top & 31,403 &  0.38 & 52.214 \\
    Four-top & 874,355 & 10.66 & 1.875 \\
    \hline
    \textbf{TOTAL} & 8,198,428 & -- & -- \\
    \hline
    \end{tabular}
    \caption{Class distribution and balanced class weights applied to each sample in the corresponding class.}
    \label{tab:class_distribution}
    \end{table}
    
    To mitigate class imbalance (Table~\ref{tab:class_distribution}), each event was assigned a sample weight inversely proportional to its class frequency, such that all classes contribute equally to the training.
    To unify the input variable space, the following minimized approach was proposed:
    \begin{itemize}
    \item An event is represented as a collection of kinematic variables for all possible objects. For each object, we use:
    \begin{itemize}
        \item $p_T$: Transverse momentum
        \item $\eta$: Pseudorapidity  
        \item $\phi$: Azimuthal angle
        \item $P_x, P_y, P_z$: Cartesian momentum components
    \end{itemize}
    \item Event-wide variables include:
    \begin{itemize}
        \item $N_{\text{jets}}$: Number of jets
        \item $N_{b\text{-jets}}$: Number of b-tagged jets
        \item $N_\nu$: Number of neutrinos
        \item $N_l$: Number of leptons
        \item $\text{MET}$: Missing Transverse Energy
    \end{itemize}
    \item Maximum object counts: 12 jets (4 b-jets) + 4 leptons.
    \item Objects within each group are sorted by energy. 
    \item Missing objects are represented with zero-valued variables, handled separately.
    \end{itemize}
    \section{Event representation learning}
    
    \begin{figure}[h]
        \centering
        \hspace*{\fill}
        \includegraphics[width=0.49\textwidth]{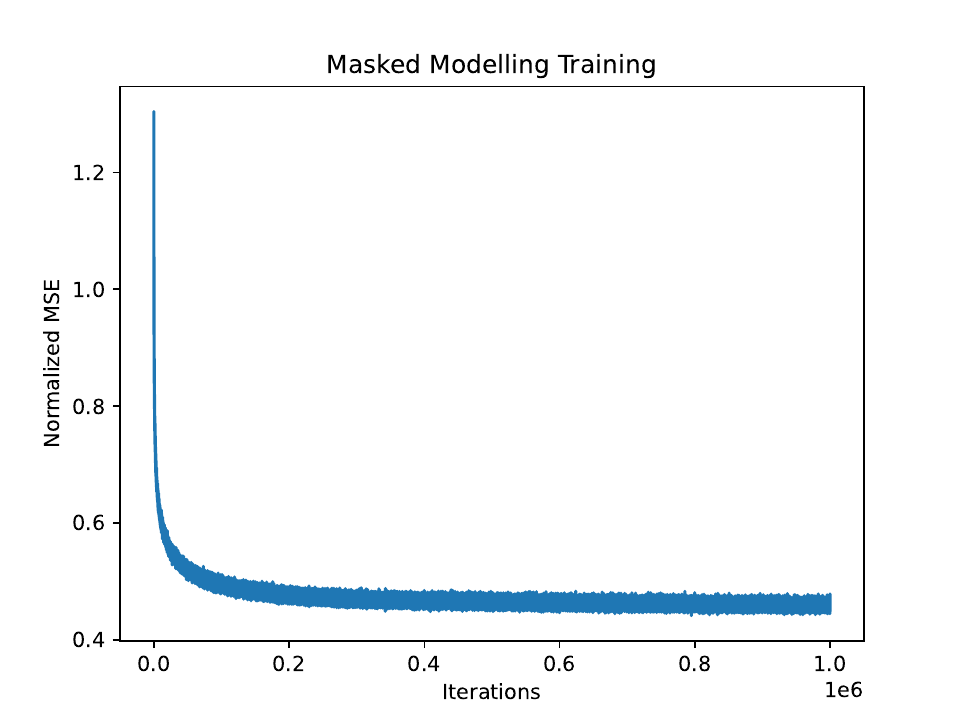} \hfill
        \includegraphics[width=0.49\textwidth]{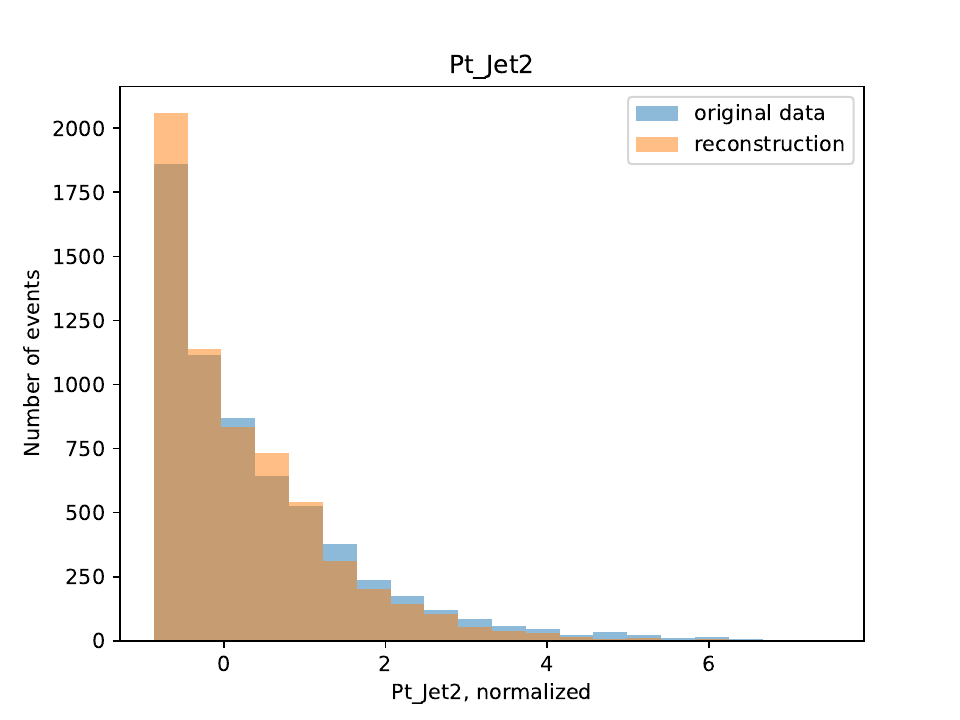}
        \hspace*{\fill} \\[1em]
        
        \hspace*{\fill}
        \includegraphics[width=0.49\textwidth]{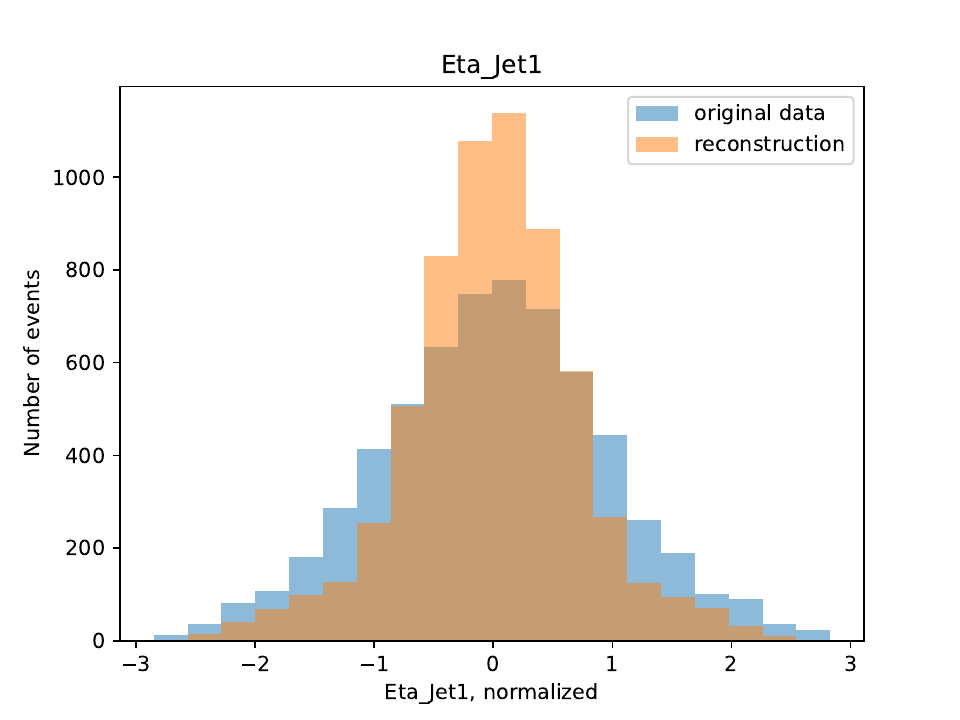} \hfill
        \includegraphics[width=0.49\textwidth]{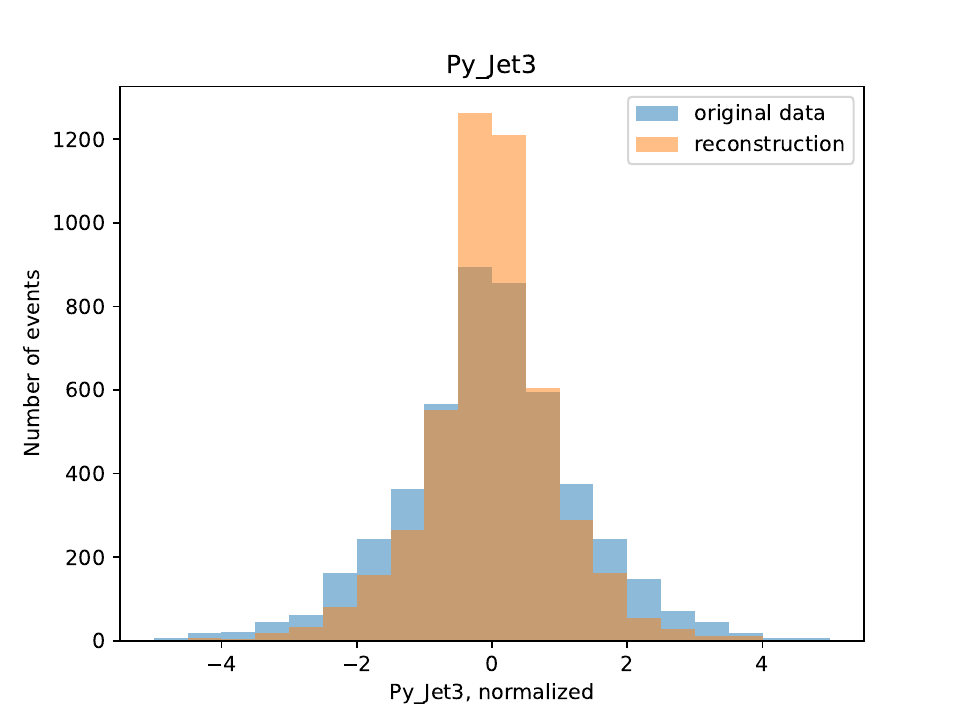}
        \hspace*{\fill}
        
        \caption{\label{fig:masked_modelling} Examples of pre-training results for masked reconstruction (average reconstruction error for the test split and event distributions for some of the reconstructed variables). Masking probability: 30\%, standardized (i.e. subtracted the mean and divided by the standard deviation) variables.}
        \label{fig:masked_modelling}
    \end{figure}
    
    A potential step for analyzing processes with different kinematic features involves creating a unified representation space. Typically, this is the output of an intermediate network layer trained on a combined dataset. This space is expected to better reveal data patterns and cluster definitions. Common implementations use dimensionality reduction (autoencoders) or masked variable modeling~\cite{Golling:2024abg}.
    
\begin{figure}[H]
    \centering
    
    \begin{minipage}{\textwidth}
        \centering
        \includegraphics[width=0.49\textwidth]{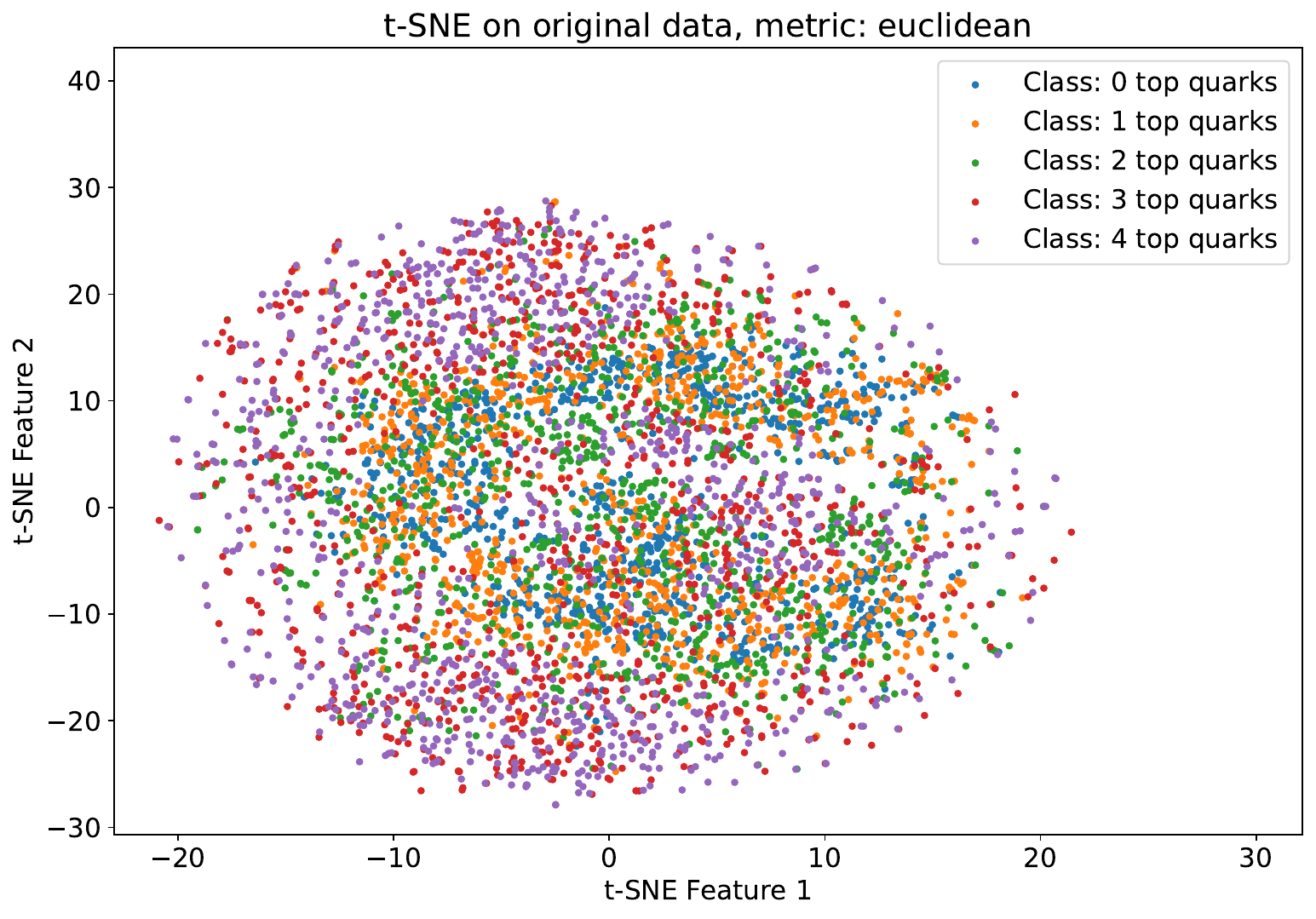}%
        \hfill
        \includegraphics[width=0.49\textwidth]{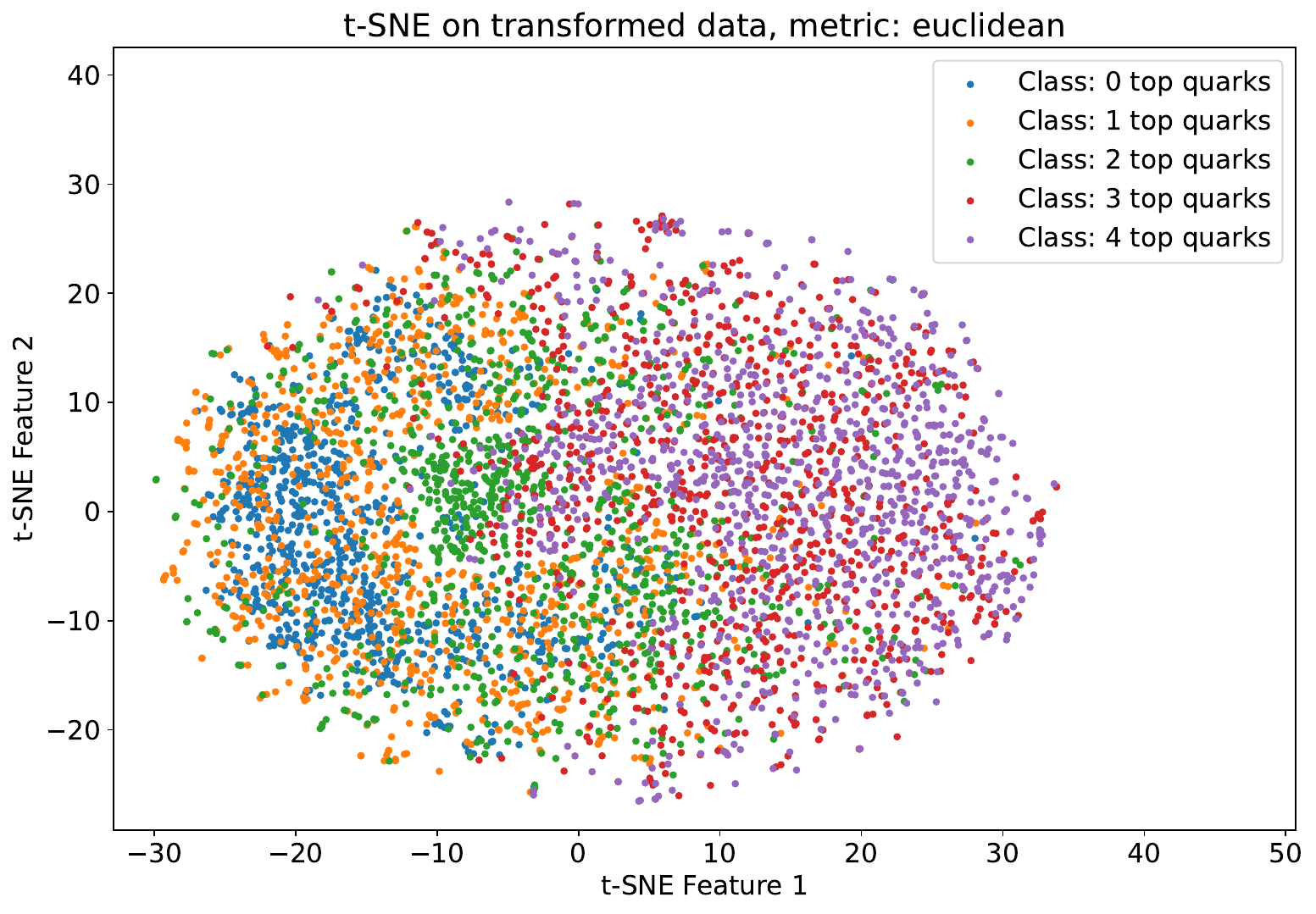}
        \par\vspace{0.5em}
        \small\textbf{(a)} Euclidean distance: original input (left) vs. transformed space (right).
    \end{minipage}
    
    \vspace{1em} 
    
    \begin{minipage}{\textwidth}
        \centering
        \includegraphics[width=0.49\textwidth]{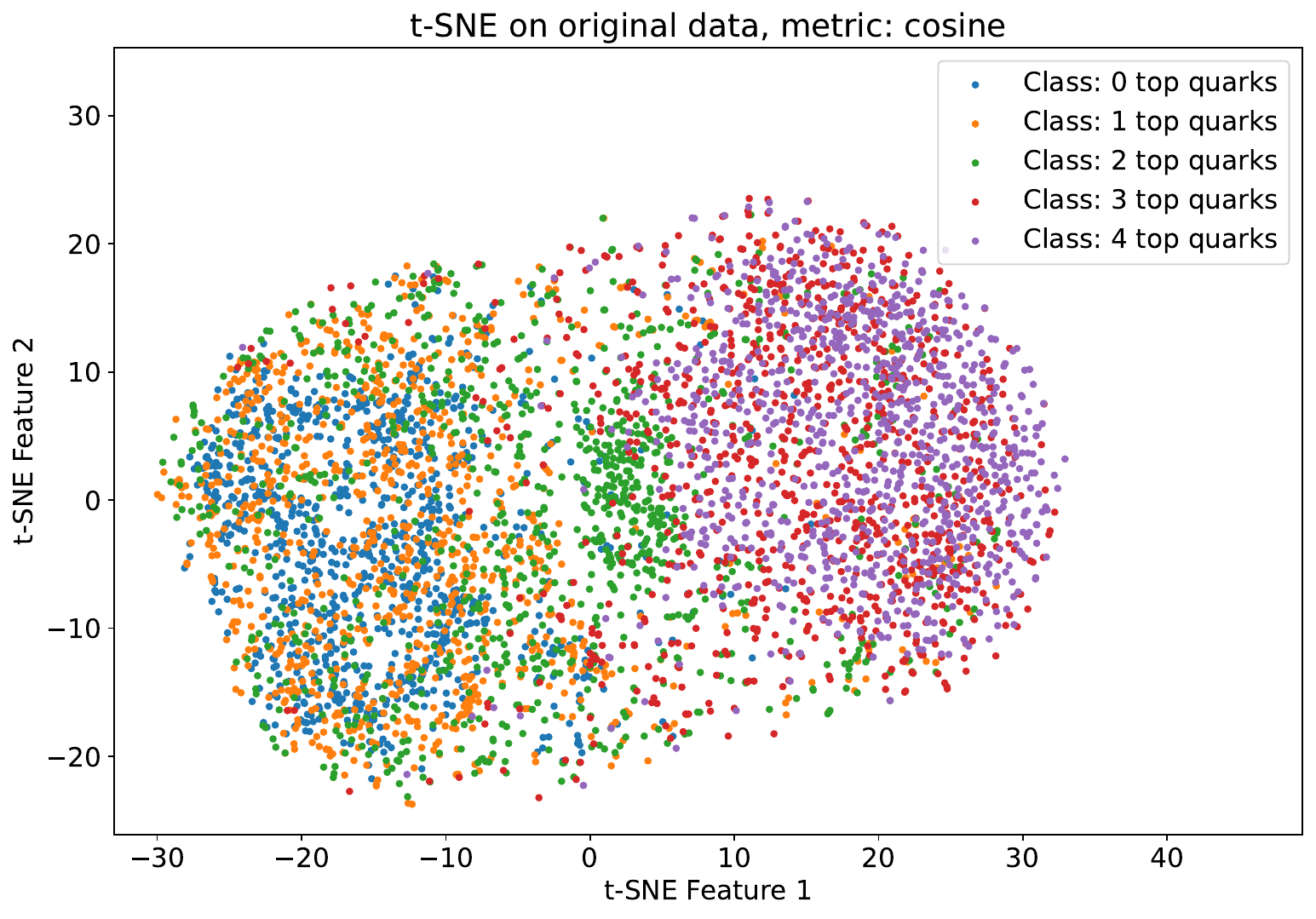}%
        \hfill
        \includegraphics[width=0.49\textwidth]{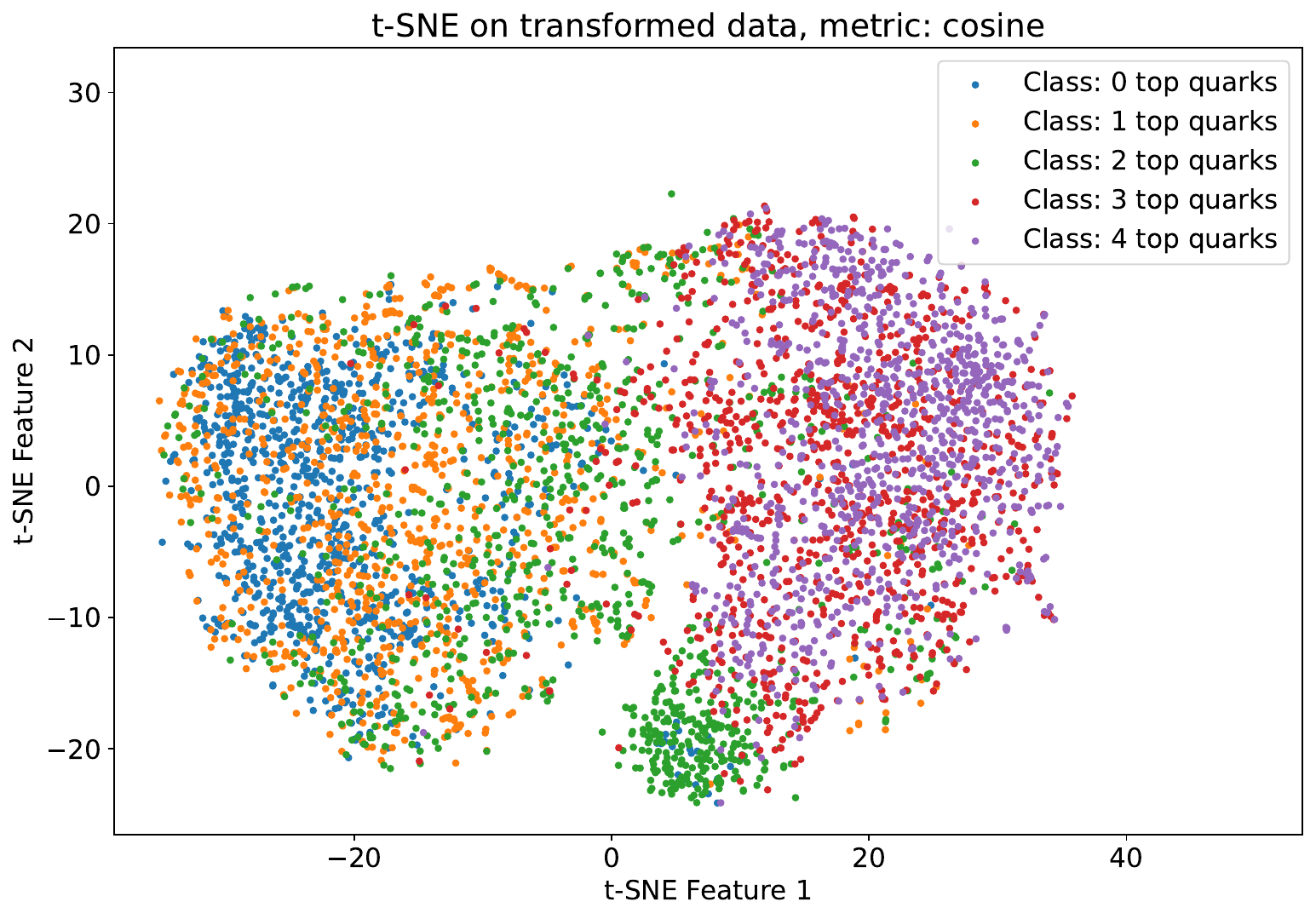}
        \par\vspace{0.5em}
        \small\textbf{(b)} Cosine distance: original input (left) vs. transformed space (right).
    \end{minipage}
    
    \vspace{1em}
    
    \begin{minipage}{\textwidth}
        \centering
        \includegraphics[width=0.49\textwidth]{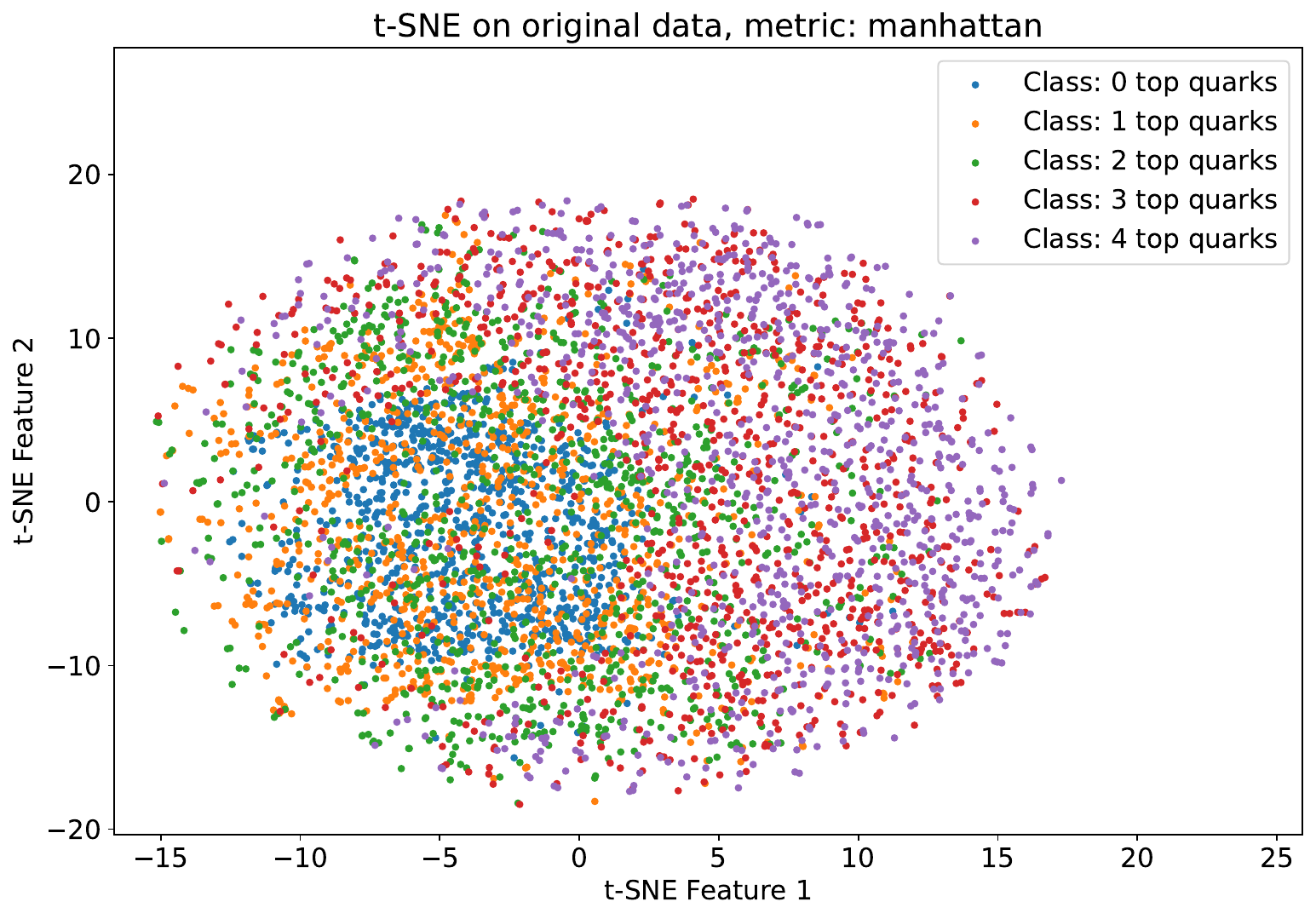}%
        \hfill
        \includegraphics[width=0.49\textwidth]{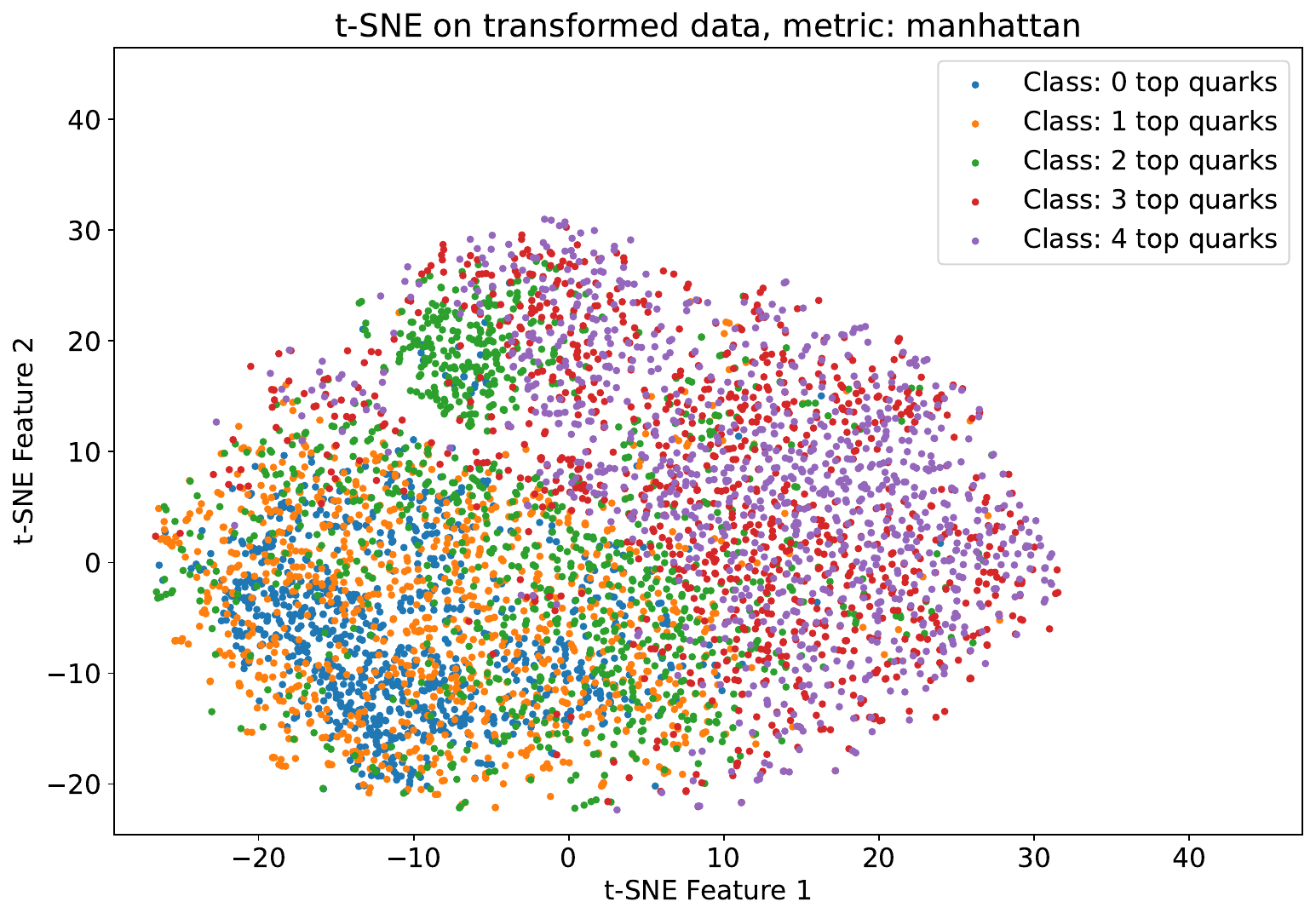}
        \par\vspace{0.5em}
        \small\textbf{(c)} Manhattan distance: original input (left) vs. transformed space (right).
    \end{minipage}

    \caption{t-SNE~\cite{JMLR:v9:vandermaaten08a} visualization of the effect of DNN transformation across distance metrics. 
    Each row compares original input space (left) with transformed space (right).}
    \label{fig:masked_modelling_tsne}
\end{figure}

    For our task, masked variable reconstruction was chosen (Fig.~\ref{fig:masked_modelling}). Compared to autoencoders, this method may improve precision by masking all variables during training, forcing the model to learn inter-variable relationships rather than relying on average reconstruction quality. The visualization of the data manifold before and after applying the masked modelling approach are presented in (Fig.~\ref{fig:masked_modelling_tsne}). The clusters containing events with the same class are more pronounced after applying the model. 
    
    Potential pre-training directions include using generator-level information: neutrino component regression and combinatorial challenges, with unfolding to parton-level information.

    
        
    
    

    \section{Transformer training}
    
\begin{figure}[H]
    \centering
    \includegraphics[width=0.45\textwidth]{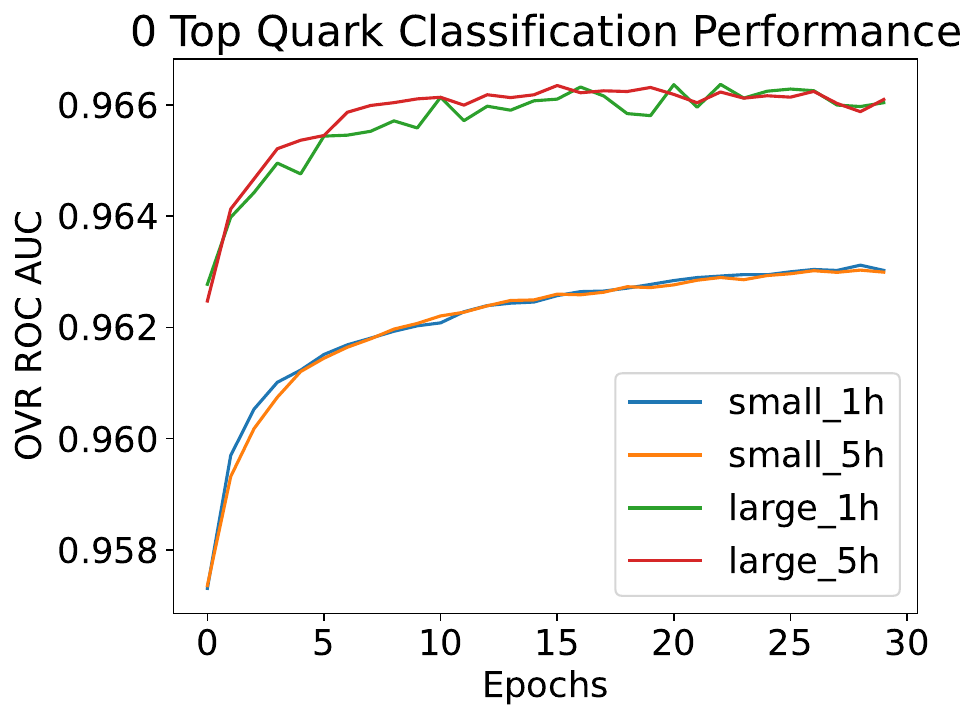} \hfill
    \includegraphics[width=0.45\textwidth]{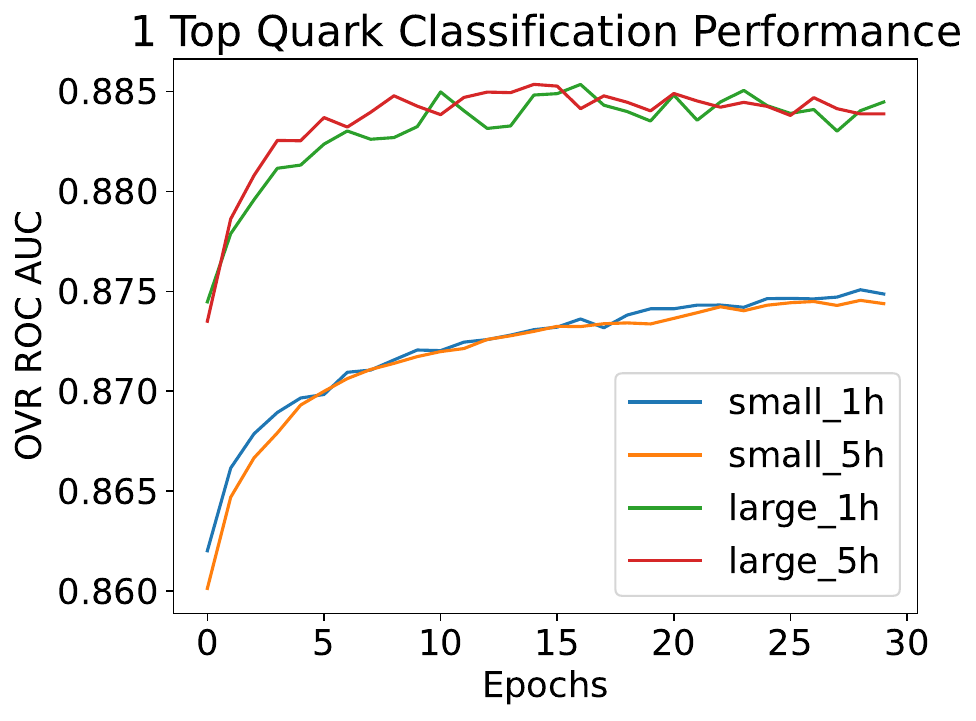} \\[1em]
    
    \includegraphics[width=0.45\textwidth]{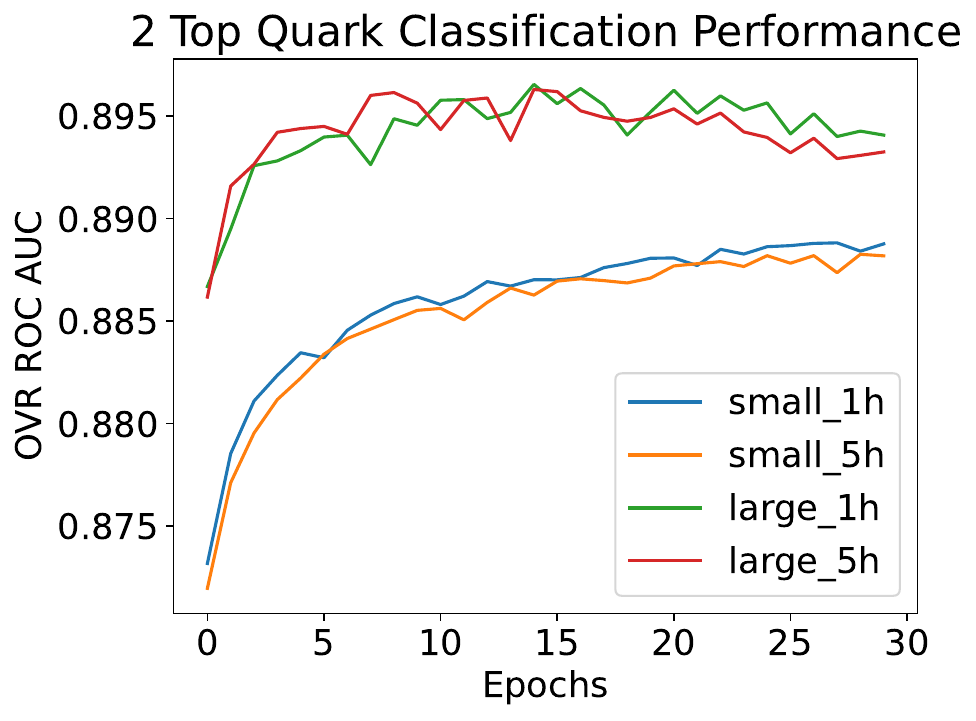} \hfill
    \includegraphics[width=0.45\textwidth]{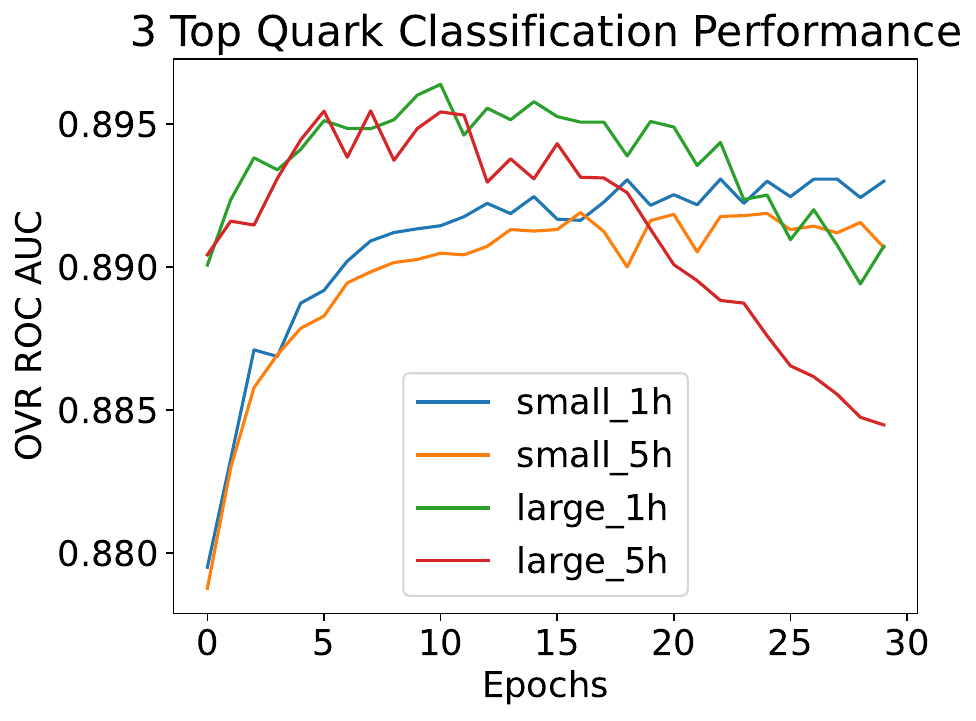} \\[1em]
    
    \includegraphics[width=0.45\textwidth]{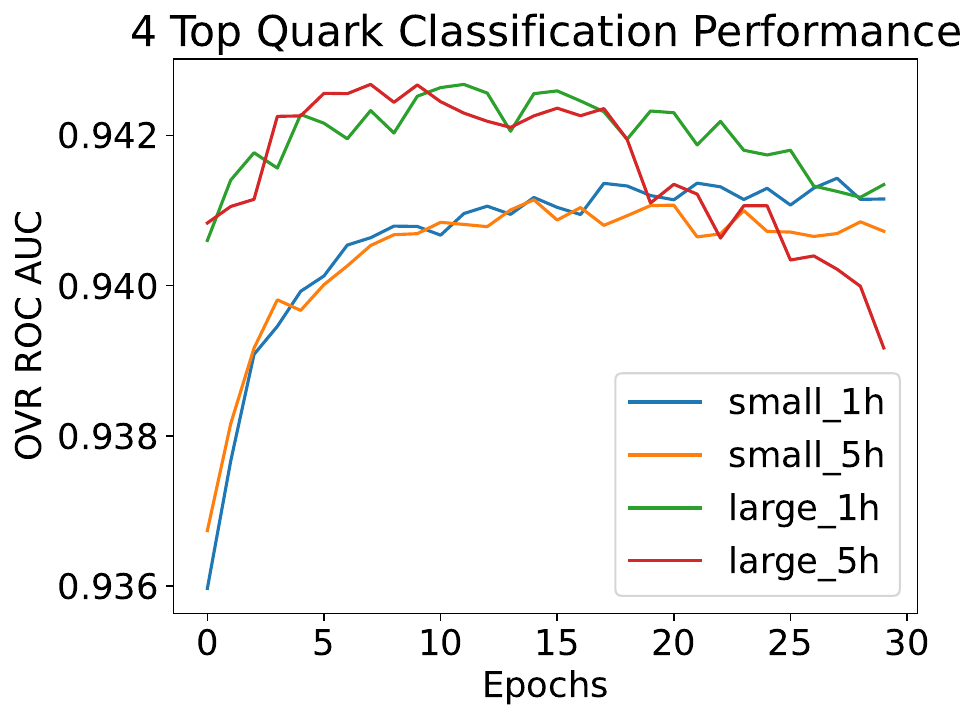}
    
    \caption{One-vs-rest ROC AUC metrics for all classes and reported architectures.}
    \label{fig:roc_auc_metrics}
\end{figure}
    
    To study architectural dependencies, four Transformer network variants~\cite{Attention} were prepared (Table~\ref{tab:hyperparams}). Preliminary comparisons (Fig.~\ref{fig:roc_auc_metrics}) revealed:
    \begin{itemize}
    \item Model size increases improve classification metrics but require more training resources
    \item Adding attention heads degrades performance when overfitting occurs
    \end{itemize}
    \begin{table}[H]
    \centering
    \caption{Transformer architecture variants and training hyperparameters. All models use AdamW optimizer with learning rate 0.0001 and no weight decay. The model metrics are calculated as the maximum mean one-vs-rest Receiver Operating Characteristic Area Under the Curve (ROC AUC) on the test set.}
    \label{tab:hyperparams}
    \begin{tabular}{lcccc}
    \toprule
    Parameter & Small\_1h & Small\_5h & Large\_1h & Large\_5h \\
    \midrule
    Input size & 125 & 125 & 125 & 125 \\
    Output size & 5 & 5 & 5 & 5 \\
    Feedforward dim & 128 & 128 & 512 & 512 \\
    Hidden layers & 5 & 5 & 5 & 5 \\
    Embedding dim & 16 & 20 & 125 & 125 \\
    Attention heads & 1 & 5 & 1 & 5 \\
    Dropout & 0.1 & 0.1 & 0.1 & 0.1 \\
    \midrule
    Max of Mean ROC AUC & 0.9128 & 0.9121 & 0.9172 & 0.9169 \\
    \bottomrule
    \end{tabular}
    \end{table}
    
    \section{Example of downstream application}
    
\begin{figure}[H]
    \centering

    \begin{subfigure}{0.9\textwidth}
        \centering
        \begin{minipage}{0.45\textwidth}
            \centering
            \includegraphics[width=80mm]{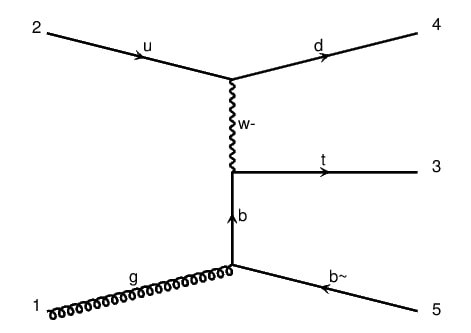}
        \end{minipage}
        \hfill
        \begin{minipage}{0.45\textwidth}
            \centering
            \includegraphics[width=80mm]{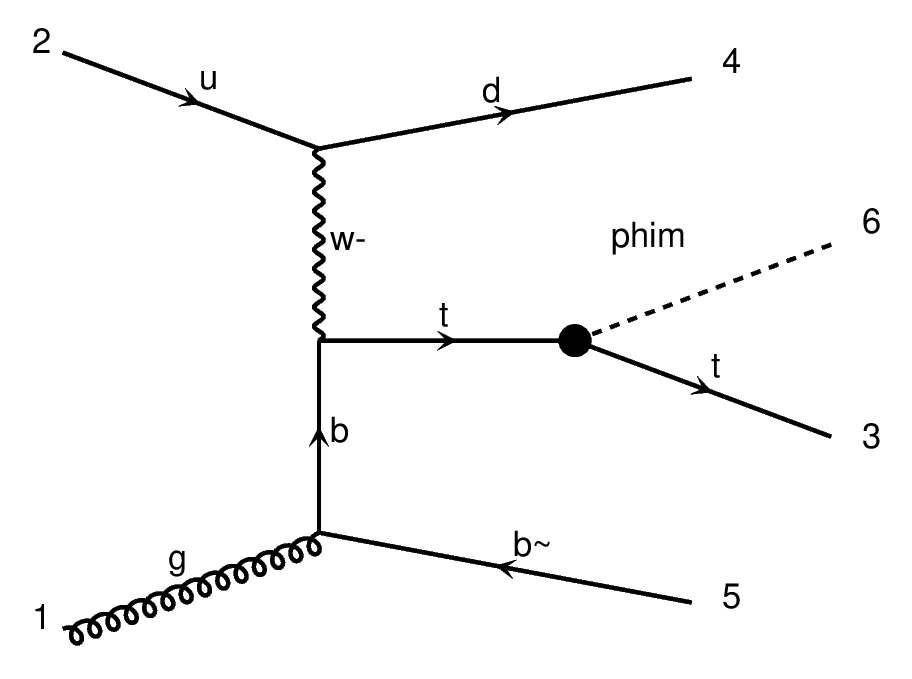}
        \end{minipage}
        \caption{Examples of Feynman diagrams for background (SM, left) and signal (DM, right) processes.}
        \label{fig:diagrams}
    \end{subfigure}

    \vspace{1em}
    \begin{subfigure}{0.9\textwidth}
        \centering
        \includegraphics[width=160mm]{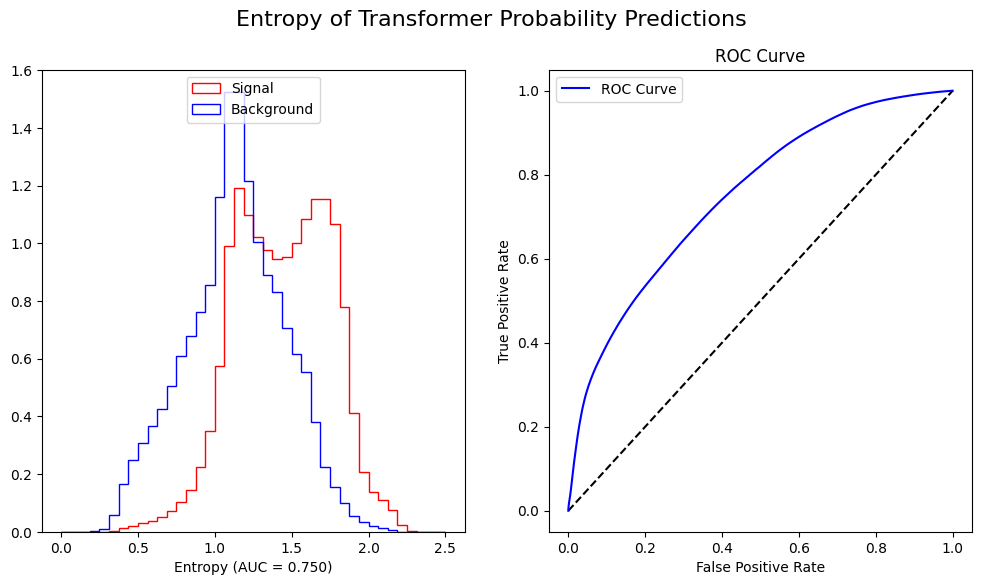}
        \caption{Separating power of entropy distributions for background vs. signal events.}
        \label{fig:entropy}
    \end{subfigure}

    \caption{\label{fig:dm_entropy} 
    Feynman diagrams and entropy-based separation of background and signal.}
\end{figure}

    As an application example, the entropy of the Transformer's probability distribution (without fine-tuning) was used to identify events with dark matter (DM) mediator in single top-quark production (Fig.~\ref{fig:dm_entropy}). The specifics of the DM model are covered in~\cite{Abasov:2025ntj, Bunichev:2024}. It is important to emphasize that the Transformer was trained exclusively on the Standard Model events and was not exposed to DM events during training. The entropy function \(H(X) = -\sum_{x \in X} p(x) \log_{2} p(x)\) was applied to predicted probabilities for each event, comparing t-channel single top-quark production in Standard Model (SM) and in association with scalar dark matter mediator in Simplified Model scenario (DM).
    
    \section{Prospects}
    Future work will focus on building a universal model for different processes with top-quarks. One of the most promising applications is the study of 3- and 4-top-quarks production processes for dark matter searches~\cite{Craig:2016ygr, ATLAS:2022rws}. To enhance sensitivity to these rare processes, foundational model pre-training will be implemented. Training efficient models on diverse datasets could reveal complex patterns potentially boosting specialized analysis sensitivity in all possible processes with top-quark in different theories. Such foundational NN models already show promise in HEP~\cite{ATLAS:2024gth, Qu:2022mxj}, demonstrating potential for unifying analyses of disparate processes.
    
    Some additional studies could include incorporating various methods implemented in state-of-the-art neural networks for particle reconstructions and unfolding. One approach is utilizing the tensor attention mechanism~\cite{Shmakov:2021qdz}, which includes symmetry between decay products in both the network architecture and loss function. Since the data has azimuthal symmetry, the covariant attention mechanism~\cite{Li:2022xfc, Qiu:2022xvr}, which uses Lorentz symmetry to account for this fact, also looks promising; however, studies conducted in~\cite{Li:2022xfc, Qiu:2022xvr, Fenton:2023ikr} show that while this approach improves network performance if trained on small datasets ($< 10^5$ events), its effect becomes negligible as the size of the dataset increases. Even in the current state of our research, we use approximately $5*10^6$ events for training, so the traditional attention mechanism will provide sufficient performance.
    
    At this stage of the work, only basic kinematic variables are used as inputs for the network. This standardized set is used in almost all state-of-the-art event reconstruction methods, such as SPA-NET~\cite{Fenton:2023ikr} and HyPER~\cite{Birch-Sykes:2024gij} and has proven to be sufficient for highly efficient models. However, it was demonstrated that additional variables responsible for the pairwise and cluster interaction between particles, such as scalar products, can improve performance by decreasing the possible order of nonlinearity in the functional dependence of the input space and desired output~\cite{Dudko:2020qas}. We believe that the inclusion of these variables can make a positive impact on our network and this will be studied in future research.
    
    \bibliographystyle{unsrt}
    \bibliography{main}

@article{Dudko:2020qas,
    author = "Dudko, Lev and Vorotnikov, Georgi and Volkov, Petr and Ovchinnikov, Dmitri and Perfilov, Maxim and Shporin, Artem and Chernoded, Andrei",
    title = "{General recipe to form input space for deep learning analysis of HEP scattering processes}",
    eprint = "2002.09350",
    archivePrefix = "arXiv",
    primaryClass = "hep-ph",
    doi = "10.1142/S0217751X20501195",
    journal = "Int. J. Mod. Phys. A",
    volume = "35",
    number = "21",
    pages = "2050119",
    year = "2020"
}

@article{ATLAS:2022rws,
    author = "Aad, Georges and others",
    collaboration = "ATLAS",
    title = "{Search for $ t\overline{t}H/A\to t\overline{t}t\overline{t} $ production in the multilepton final state in proton\textendash{}proton collisions at $ \sqrt{s} $ = 13 TeV with the ATLAS detector}",
    eprint = "2211.01136",
    archivePrefix = "arXiv",
    primaryClass = "hep-ex",
    reportNumber = "CERN-EP-2022-170",
    doi = "10.1007/JHEP07(2023)203",
    journal = "JHEP",
    volume = "07",
    pages = "203",
    year = "2023"
}

@article{Craig:2016ygr,
    author = "Craig, Nathaniel and Hajer, Jan and Li, Ying-Ying and Liu, Tao and Zhang, Hao",
    title = "{Heavy Higgs bosons at low $\tan \beta$: from the LHC to 100 TeV}",
    eprint = "1605.08744",
    archivePrefix = "arXiv",
    primaryClass = "hep-ph",
    doi = "10.1007/JHEP01(2017)018",
    journal = "JHEP",
    volume = "01",
    pages = "018",
    year = "2017"
}

@article{ATLAS:2024gth,
    author = "Aad, Georges and others",
    collaboration = "ATLAS",
    title = "{Measurement of the associated production of a top-antitop-quark pair and a Higgs boson decaying into a $b\bar{b}$ pair in pp collisions at $\sqrt{s}=13$~TeV using the ATLAS detector at the LHC}",
    eprint = "2407.10904",
    archivePrefix = "arXiv",
    primaryClass = "hep-ex",
    reportNumber = "CERN-EP-2024-194",
    doi = "10.1140/epjc/s10052-025-13740-x",
    journal = "Eur. Phys. J. C",
    volume = "85",
    number = "2",
    pages = "210",
    year = "2025"
}

@InProceedings{Qu:2022mxj,
    author = "Qu, Huilin and Li, Congqiao and Qian, Sitian",
    title = "{Particle Transformer} for Jet Tagging",
    booktitle = "{Proceedings of the 39th International Conference on Machine Learning}",
    pages = "18281--18292",
    year = "2022",
    eprint = "2202.03772",
    archivePrefix = "arXiv",
    primaryClass = "hep-ph",
    note = "\texttt{arXiv:2202.03772 [hep-ph]}"
}

@article{Alwall:2014hca,
    author = "Alwall, J. and Frederix, R. and Frixione, S. and Hirschi, V. and Maltoni, F. and Mattelaer, O. and Shao, H. -S. and Stelzer, T. and Torrielli, P. and Zaro, M.",
    title = "{The automated computation of tree-level and next-to-leading order differential cross sections, and their matching to parton shower simulations}",
    eprint = "1405.0301",
    archivePrefix = "arXiv",
    primaryClass = "hep-ph",
    reportNumber = "CERN-PH-TH-2014-064, CP3-14-18, LPN14-066, MCNET-14-09, ZU-TH-14-14",
    doi = "10.1007/JHEP07(2014)079",
    journal = "JHEP",
    volume = "07",
    pages = "079",
    year = "2014"
}

@article{Selvaggi:2014mya,
    author = "Selvaggi, Michele",
    editor = "Wang, Jianxiong",
    title = "{DELPHES 3: A modular framework for fast-simulation of generic collider experiments}",
    doi = "10.1088/1742-6596/523/1/012033",
    journal = "J. Phys. Conf. Ser.",
    volume = "523",
    pages = "012033",
    year = "2014"
}

@article{CompHEP:2004qpa,
    author = "Boos, E. and Bunichev, V. and Dubinin, M. and Dudko, L. and Ilyin, V. and Kryukov, A. and Edneral, V. and Savrin, V. and Semenov, A. and Sherstnev, A.",
    editor = "Kawabata, S. and Perret-Gallix, D.",
    collaboration = "CompHEP",
    title = "{CompHEP 4.4: Automatic computations from Lagrangians to events}",
    eprint = "hep-ph/0403113",
    archivePrefix = "arXiv",
    doi = "10.1016/j.nima.2004.07.096",
    journal = "Nucl. Instrum. Meth. A",
    volume = "534",
    pages = "250--259",
    year = "2004"
}

@article{Pukhov:1999gg,
    author = "Pukhov, A. and Boos, E. and Dubinin, M. and Edneral, V. and Ilyin, V. and Kovalenko, D. and Kryukov, A. and Savrin, V. and Shichanin, S. and Semenov, A.",
    title = "{CompHEP: A Package for evaluation of Feynman diagrams and integration over multiparticle phase space}",
    eprint = "hep-ph/9908288",
    archivePrefix = "arXiv",
    reportNumber = "INP-MSU-98-41-542",
    year = "1999",
    note = "\texttt{arXiv:hep-ph/9908288}"
}

@article{Golling:2024abg,
    author = "Golling, Tobias and Heinrich, Lukas and Kagan, Michael and Klein, Samuel and Leigh, Matthew and Osadchy, Margarita and Raine, John Andrew",
    title = "{Masked particle modeling on sets: towards self-supervised high energy physics foundation models}",
    eprint = "2401.13537",
    archivePrefix = "arXiv",
    primaryClass = "hep-ph",
    doi = "10.1088/2632-2153/ad64a8",
    journal = "Mach. Learn. Sci. Tech.",
    volume = "5",
    number = "3",
    pages = "035074",
    year = "2024"
}

@inproceedings{Attention,
 author = {Vaswani, Ashish and Shazeer, Noam and Parmar, Niki and Uszkoreit, Jakob and Jones, Llion and Gomez, Aidan N and Kaiser, \L ukasz and Polosukhin, Illia},
 booktitle = {Advances in Neural Information Processing Systems},
 editor = {I. Guyon and U. Von Luxburg and S. Bengio and H. Wallach and R. Fergus and S. Vishwanathan and R. Garnett},
 pages = {},
 publisher = {Curran Associates, Inc.},
 title = {Attention is All you Need},
 url = {https://proceedings.neurips.cc/paper_files/paper/2017/file/3f5ee243547dee91fbd053c1c4a845aa-Paper.pdf},
 volume = {30},
 year = {2017}
}

@article{Li:2022xfc,
    author = "Li, Congqiao and Qu, Huilin and Qian, Sitian and Meng, Qi and Gong, Shiqi and Zhang, Jue and Liu, Tie-Yan and Li, Qiang",
    title = "{Does Lorentz-symmetric design boost network performance in jet physics?}",
    eprint = "2208.07814",
    archivePrefix = "arXiv",
    primaryClass = "hep-ph",
    doi = "10.1103/PhysRevD.109.056003",
    journal = "Phys. Rev. D",
    volume = "109",
    number = "5",
    pages = "056003",
    year = "2024"
}

@article{Qiu:2022xvr,
    author = "Qiu, Shikai and Han, Shuo and Ju, Xiangyang and Nachman, Benjamin and Wang, Haichen",
    title = "{Holistic approach to predicting top quark kinematic properties with the covariant particle transformer}",
    eprint = "2203.05687",
    archivePrefix = "arXiv",
    primaryClass = "hep-ph",
    doi = "10.1103/PhysRevD.107.114029",
    journal = "Phys. Rev. D",
    volume = "107",
    number = "11",
    pages = "114029",
    year = "2023"
}

@article{Fenton:2023ikr,
    author = "Fenton, Michael James and Shmakov, Alexander and Okawa, Hideki and Li, Yuji and Hsiao, Ko-Yang and Hsu, Shih-Chieh and Whiteson, Daniel and Baldi, Pierre",
    title = "{Reconstruction of unstable heavy particles using deep symmetry-preserving attention networks}",
    eprint = "2309.01886",
    archivePrefix = "arXiv",
    primaryClass = "hep-ex",
    doi = "10.1038/s42005-024-01627-4",
    journal = "Commun. Phys.",
    volume = "7",
    number = "1",
    pages = "139",
    year = "2024"
}

@article{Shmakov:2021qdz,
    author = "Shmakov, Alexander and Fenton, Michael James and Ho, Ta-Wei and Hsu, Shih-Chieh and Whiteson, Daniel and Baldi, Pierre",
    title = "{SPANet: Generalized permutationless set assignment for particle physics using symmetry preserving attention}",
    eprint = "2106.03898",
    archivePrefix = "arXiv",
    primaryClass = "hep-ex",
    doi = "10.21468/SciPostPhys.12.5.178",
    journal = "SciPost Phys.",
    volume = "12",
    number = "5",
    pages = "178",
    year = "2022"
}

@article{Birch-Sykes:2024gij,
    author = "Birch-Sykes, Callum and Le, Brian and Peters, Yvonne and Simpson, Ethan and Zhang, Zihan",
    title = "{Reconstructing short-lived particles using hypergraph representation learning}",
    eprint = "2402.10149",
    archivePrefix = "arXiv",
    primaryClass = "hep-ph",
    doi = "10.1103/PhysRevD.111.032004",
    journal = "Phys. Rev. D",
    volume = "111",
    number = "3",
    pages = "032004",
    year = "2025"
}

@article{Abasov:2025ntj,
    author = "Abasov, E. and Dudko, L. and Iudin, E. and Markina, A. and Volkov, P. and Vorotnikov, G. and Perfilov, M. and Zaborenko, A.",
    title = "{Reconstruction of angular correlations in the associated top quark and the dark matter mediator production}",
    journal = "ArXiv",
    eprint = "2504.14303",
    archivePrefix = "arXiv",
    primaryClass = "hep-ph",
    month = "4",
    year = "2025",
    note = "\texttt{arXiv:2504.14303 [hep-ph]}"
}

@article{Bunichev:2024,
    author = "Boos, E. and Bunichev, V. and Dudko, L.",
    title = "{Dark Matter Mediators in Processes of Single Top Quark Production}",
    doi = "10.1134/S1063779624701740",
    journal = "Physics of Particles and Nuclei",
    volume = "56",
    number = "2",
    pages = "429-433",
    year = "2025"
}

@Article{transformer_nlp,
AUTHOR = {Patwardhan, Narendra and Marrone, Stefano and Sansone, Carlo},
TITLE = {Transformers in the Real World: A Survey on NLP Applications},
JOURNAL = {Information},
VOLUME = {14},
YEAR = {2023},
NUMBER = {4},
ARTICLE-NUMBER = {242},
URL = {https://www.mdpi.com/2078-2489/14/4/242},
ISSN = {2078-2489},
DOI = {10.3390/info14040242}
}

@Article{transformer_cv,
AUTHOR = {Jamil, Sonain and Jalil Piran, Md. and Kwon, Oh-Jin},
TITLE = {A Comprehensive Survey of Transformers for Computer Vision},
JOURNAL = {Drones},
VOLUME = {7},
YEAR = {2023},
NUMBER = {5},
ARTICLE-NUMBER = {287},
URL = {https://www.mdpi.com/2504-446X/7/5/287},
ISSN = {2504-446X},
DOI = {10.3390/drones7050287}
}

@inproceedings{gorishniy2022embeddings,
    title={On Embeddings for Numerical Features in Tabular Deep Learning},
    author={Yury Gorishniy and Ivan Rubachev and Artem Babenko},
    booktitle={{NeurIPS}},
    year={2022},
}

@article{JMLR:v9:vandermaaten08a,
  author  = {Laurens van der Maaten and Geoffrey Hinton},
  title   = {Visualizing Data using t-SNE},
  journal = {Journal of Machine Learning Research},
  year    = {2008},
  volume  = {9},
  number  = {86},
  pages   = {2579--2605},
  url     = {http://jmlr.org/papers/v9/vandermaaten08a.html}
}

@article{CMS_detector,
    author = "Chatrchyan, S. and others",
    collaboration = "CMS",
    title = "{The CMS Experiment at the CERN LHC}",
    doi = "10.1088/1748-0221/3/08/S08004",
    journal = "JINST",
    volume = "3",
    pages = "S08004",
    year = "2008"
}

@misc{CMS_card,
  title = "{DELPHES CMS Card}",
  howpublished = {\url{https://github.com/delphes/delphes/blob/master/cards/delphes_card_CMS.tcl}},
  note = {Accessed: 2025-10-02}
}
    \end{document}